\documentclass[fleqn,usenatbib]{mnras}

\usepackage{newtxtext,newtxmath}

\usepackage[T1]{fontenc}
\usepackage{ae,aecompl}

\usepackage{graphicx}	
\usepackage{amsmath}	
\usepackage{amssymb}	

\newcommand{\bb}{\langle B \rangle}     
\newcommand{\logg}{log $g$}
\newcommand{\vsini}{$v_\textrm{e} \sin i$}
\newcommand{\vmic}{$v_\text{mic}$}
\newcommand{\vmac}{$v_\text{mac}$}
\newcommand{\kms}{$\text{km } \text{s}^{-1}$}
\newcommand{\teff}{$T_{\text{eff}}$}	

\title[Magnetic field of the PMS binary V1878 Ori]{The large-scale magnetic field of the eccentric pre-main-sequence binary system V1878 Ori}

\author[A. Lavail et al.]{
A. Lavail$^{1}$%
\thanks{E-mail: alexis.lavail@physics.uu.se},
O. Kochukhov$^{1}$,
G.A.J. Hussain$^{2}$,
C. Argiroffi$^{3,4}$,
E. Alecian$^{5}$,\newauthor
J. Morin$^{6}$,
and the BinaMIcS collaboration
\\
$^{1}$Department of Physics and Astronomy, Uppsala University, Box 516, SE-751 20 Uppsala, Sweden\\
${^2}$European Southern Observatory, Karl-Schwarzschild-Strasse 2, 85748 Garching, Germany
\\
${^3}$University of Palermo, Department of Physics and Chemistry, Piazza del Parlamento 1, 90134, Palermo, Italy
\\
${^4}$INAF - Osservatorio Astronomico di Palermo, Piazza del Parlamento 1, 90134, Palermo, Italy
\\
${^5}$Universit\'e Grenoble Alpes, CNRS, IPAG, 38000 Grenoble, France\\
${^6}$LUPM, Universit\'e de Montpellier, CNRS, Place Eug\`ene Bataillon, 34095 Montpellier, France}

\date{Accepted XXX. Received YYY; in original form ZZZ}

\pubyear{2020}

\begin{document}
\label{firstpage}
\pagerange{\pageref{firstpage}--\pageref{lastpage}}
\maketitle

\begin{abstract}
We report time-resolved, high-resolution optical spectropolarimetric observations of the young double-lined spectroscopic binary V1878 Ori. Our observations were collected with the ESPaDOnS spectropolarimeter at the Canada--France--Hawaii Telescope through the BinaMIcS large programme. V1878 Ori A and B are partially convective intermediate mass weak-line T Tauri stars on an eccentric and asynchronous orbit. We also acquired X-ray observations at periastron and outside periastron. 

Using the least-squares deconvolution technique (LSD) to combine information from many spectral lines, we clearly detected circular polarization signals in both components throughout the orbit. We refined the orbital solution for the system and obtained disentangled spectra for the primary and secondary components. The disentangled spectra were then employed to determine atmospheric parameters of the two components using spectrum synthesis.

Applying our Zeeman Doppler imaging code to composite Stokes $IV$ LSD profiles, we reconstructed brightness maps and the global magnetic field topologies of the two components. We find that V1878 Ori A and B have strikingly different global magnetic field topologies and mean field strengths. The global magnetic field of the primary is predominantly poloidal and non-axisymmetric (with a mean field strength of 180~G). while the secondary has a mostly toroidal and axisymmetric global field (mean strength of 310~G). These findings confirm that stars with very similar parameters can exhibit radically different global magnetic field characteristics. The analysis of the X-ray data shows no sign of enhanced activity at periastron, suggesting the lack of strong magnetospheric interaction at this epoch.
\end{abstract}




 
\section{Introduction}
\label{section:introduction}

Magnetic fields play very important roles throughout the entire stellar evolution. They directly influence many physical processes, both within stars and in their immediate surroundings. During the early stages of stellar formation and on the pre-main-sequence (PMS), magnetic fields are particularly important, influencing for instance the collapse of molecular clouds \citep{hennebelle2019FrASS...6....5H}, the accretion mechanism \citep{hartmann2016ARA&A..54..135H}, the formation and collimation of jets and outflows \citep{pudritz2019FrASS...6...54P}, and stellar spin down \citep{bouvier2013EAS....62..143B}. The development of dedicated instrumentation, such as high-resolution spectropolarimeters, and tomographic surface mapping techniques has allowed to measure and map surface magnetic fields in a plethora of cool and hot stars \citep{donati2009ARA&A..47..333D,reiners2012LRSP....9....1R}. Yet, we do not have a complete theory that can consistently reproduce the stellar magnetic fields characteristics across the Hertzsprung--Russell diagram \citep{brun2017LRSP...14....4B}.

A promising avenue to disentangle the influence of magnetic fields from other effects is the study of binary stars. For such systems, one can assume that the stars were formed in the same conditions, with the same material, and at the same time, which allows to separate the effects of magnetism from the initial conditions. This avenue is being explored by ``The Binarity and Magnetic Interactions in various classes of stars'' (BinaMIcS) project \citep{binamics2015}. BinaMIcs aims to study the interplay between binarity and magnetism. This project was allocated two large programmes that ran between 2013 and 2017 with the twin spectropolarimeters ESPaDOnS and Narval, respectively at the Canada-France-Hawaii-Telescope and the T\'elescope Bernard Lyot at the Pic du Midi observatory. Until now, magnetic fields of pre-main-sequence double-lined binary systems have been seldom studied through tomographic techniques such as Zeeman Doppler imaging (ZDI). The magnetic field maps of two systems, HD~155555 ($M_{1,2} \approx 1 M_\odot$) and V4046~Sgr ($M \approx 0.9 M_\odot$), were published by \citet{dunstone2008MNRAS.387.1525D} and \citet{donati2011MNRAS.417.1747D}, respectively.

V1878~Ori (= Parenago~523 = RX~J0530.7-0434 = 1RXS J053043.1-043453) is a weak-line T Tauri star binary system with an eccentric orbit and two nearly equal-mass components. The spectroscopic binary nature of V1878~Ori was first reported by \citet{alcala2000A&A...353..186A}. Shortly afterwards, the object was studied by \citet{covino2001} using time series of high-resolution optical spectra. They measured variation of the radial velocities (RVs) of each component with the cross-correlation technique. With the help of these measurements, they determined orbital parameters and reported an orbital period of $P_\text{orb} = 40.5738\pm0.0047$~days and a relatively high eccentricity with $e = 0.3181\pm 0.002$. These authors inferred nearly equal mass and luminosity for the two components, with $M_A/M_B = 1.002$ and $L_A/L_B = 1.0$, and determined that both components have a K2--K3 spectral type and a projected rational velocity {\vsini} $=13 \pm 2$~{\kms}. Photometric data from \citet{covino2001ASPC..223..503C} showed variability, possibly tracing a rotation period of 13.5~days, which amounts to about $1/3$ of the system's orbital period. Through a follow-up photometric monitoring, \citet{marilly2007} reported a photometric rotation period of 12.9 days, continuing to hint towards a non-synchronous rotation.

V1878~Ori was also observed in the radio at millimeter wavelengths \citep{kospal2011A&A...527A..96K} and in X-rays \citep{getman2016AJ....152..188G} to hunt for signs of flares or enhanced emissions at periastron where the magnetospheres could interact and magnetic reconection occur. However, no millimeter wavelength signal was detected from V1878 Ori  by \citet{kospal2011A&A...527A..96K}, and the modelling of the X-ray observations shows only a non-significant increase of the X-ray luminosity at periastron \citep{getman2016AJ....152..188G}.

According to the orbital solution by \citet{covino2001}, confirmed in this work (see Table~\ref{table:orbital-solution}), both components of V1878~Ori have a mass of at least 1.5--1.6$~M_\odot$. This means that it is the first binary system of intermediate-mass T Tauri stars (IMTTS) that can be studied with ZDI. IMTTS, with masses $1 M_\odot < M < 4 M_\odot $, are at a particularly interesting stage of stellar evolution. These stars are believed to be the precursors of PMS Herbig Ae/Be stars and A- and B-type stars on the main-sequence (MS). While all low-mass T Tauri stars ($M < 1M_\odot$) harbour a convective envelope and a priori generate a magnetic field, only 5--10\% of Herbig Ae/Be stars and A/B stars show a detectable magnetic field \citep{alecian2013,sikora2019MNRAS.483.3127S}, which has quite different characteristics than for low-mass stars. Studying magnetic fields of IMTTS is therefore a means to understand the origin of magnetism in intermediate-mass stars \citep{lavail2017A&A...608A..77L,villebrun2019A&A...622A..72V}.

Our paper is structured as follows. We present our spectropolarimetric and X-ray observations in Sect.~\ref{section:observations}. In Sect.~\ref{section:LSD} we describe our application of the least-squares deconvolution (LSD) technique to the observed data and detection of the Zeeman signatures in Stokes $V$ LSD profiles. We discuss in Sect.~\ref{section:spectral-disentangling} our spectral disentangling procedure, yielding a refined orbital solution as well as a disentangled spectra for each component. The disentangled spectra are analyzed to determine stellar parameters as described in Sect.~\ref{section:stellar-parameters}. Finally, we present in Sect.~\ref{section:zdi} simultaneous reconstruction of the global magnetic field maps for the two components using ZDI, the analysis of the X-ray data in Sect.~\ref{section:xraydata}, and discuss our results in Sect.~\ref{section:results+discussion}.

\section{Observations}
\label{section:observations}

\begin{figure}
	\includegraphics[width=1.0\columnwidth]{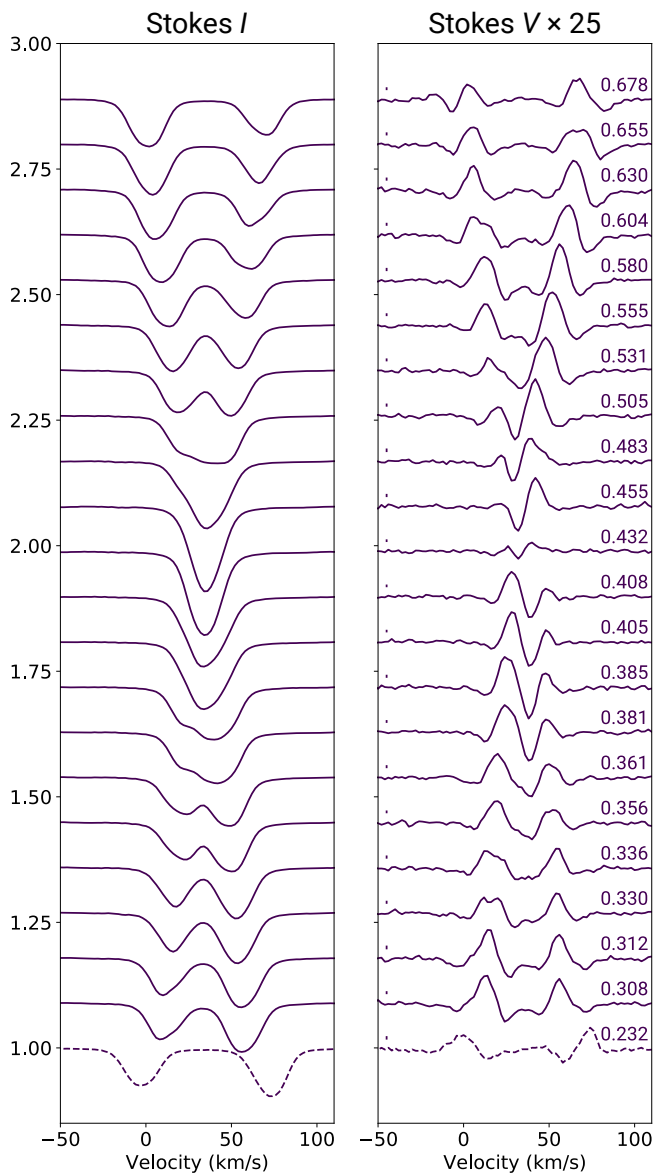}
  \caption{LSD Stokes $IV$ profiles shifted vertically with orbital phase (indicated over the Stokes $V$ profiles) increasing upwards. The observation from Feb. 2014 is indicated in dashed lines at the bottom, the rest of the observations were acquired in Jan. 2016. The median error bar for each Stokes $V$ LSD profile is plotted above each profile on the left side of the panel.}
  \label{fig:lsd}
\end{figure}

Our study of V1878~Ori is based on the high-resolution optical spectropolarimetric observations obtained in the context of the {BinaMIcS} large programme \citep{binamics2015} between 2014 and 2016. The data were acquired with the Echelle SpectroPolarimetric Device for the Observation of Stars (ESPaDOnS) instrument mounted at the 3.6-metre Canada-France-Hawaii Telescope (CFHT). ESPaDOnS \citep{donati2003-espadons,donati++2006-espadons} is a high-resolution ($R \equiv \lambda / \Delta \lambda \approx 65000$) cross-dispersed spectropolarimeter with a spectral coverage from 370 to 1000~nm that can record circularly and linearly polarized spectra as well as intensity spectra. Each polarized observation consists of a set of four sub-exposures during which the configuration of the polarimeter is changed to swap the position of the two orthogonal polarization beams through the instrument optics and on the detector. This beam-switching technique \citep{semel1993A&A...278..231S} efficiently removes instrumental polarization and effects arising from inhomogeneities between the two optical paths.

\begin{figure*}
	\includegraphics[width=1.6\columnwidth]{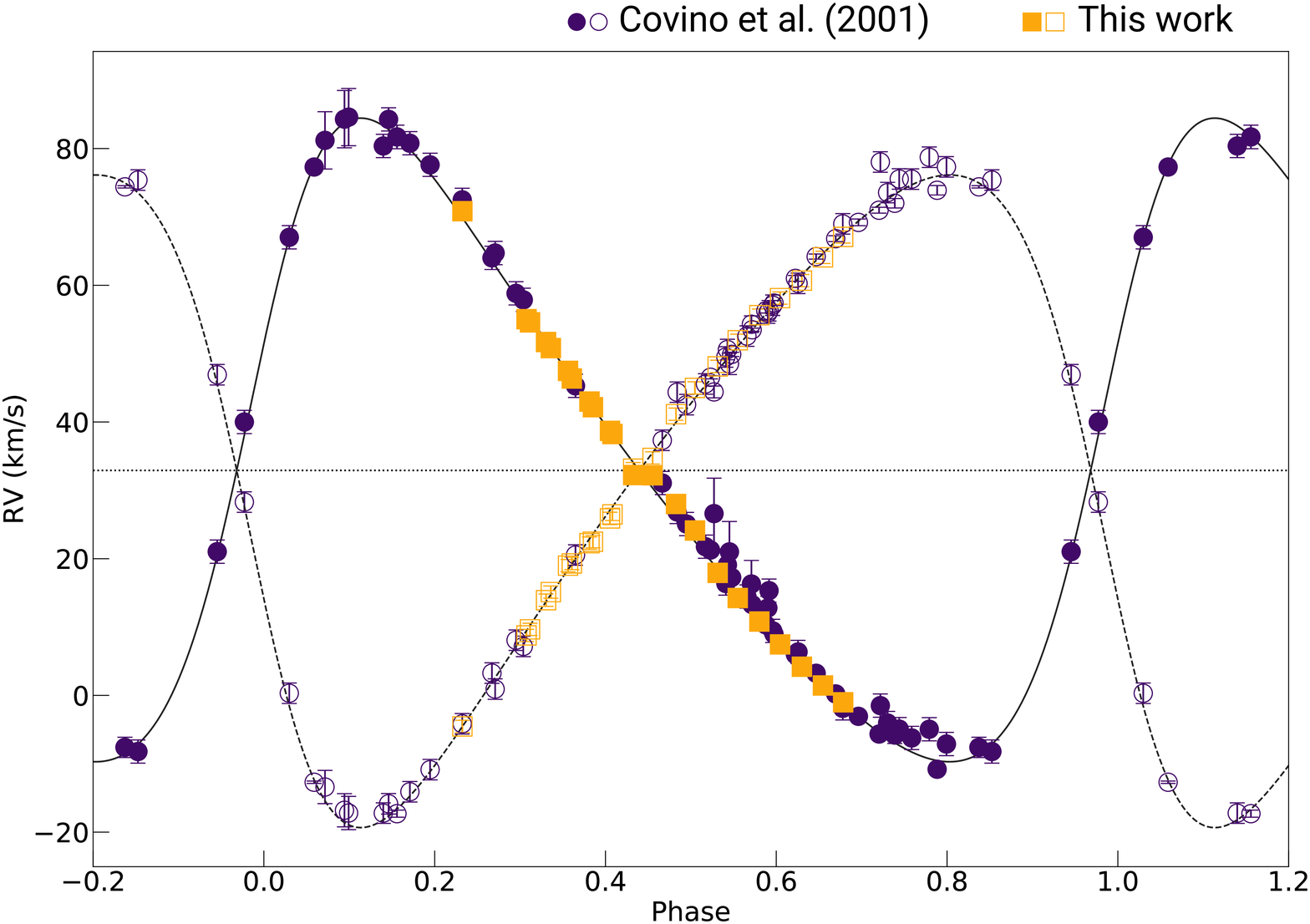}
	\caption{Radial velocity measurements for each component from \citet{covino2001} (purple circles) and this work (orange squares) and the fitted solution (black line). The primary component is indicated by filled symbols and the secondary by outlined symbols.}
  \label{fig:orbit_fit}
\end{figure*}

We obtained a total of $22$ Stokes $IV$ observations taken with exposure times of either $4 \times 1715$~s (for the observing date 2014-02-21) or $4 \times 1355$~s (all other observations). The main dataset is comprised of 21 observations obtained over 16 consecutive nights in January 2016. Our data were reduced on the fly at the telescope with the \texttt{Upena} pipeline\footnote{\url{ http://www.cfht.hawaii.edu/Instruments/Upena/}} that runs the \texttt{LIBRE-ESPRIT} routines \citep{donati++1997-libre-esprit}. The median S/N of the reduced data per spectral pixel at $\lambda = 566$~nm is $128$. Information on individual observations can be found in the observing log given in Table~\ref{table:obs}.

We complemented our study of the magnetic properties of V1878~Ori by analyzing also a long {\it XMM-Newton} observation. We observed V1878~Ori with {\it XMM-Newton} for a total of $\sim220$\,ks. The aim of the observation was to monitor its X-ray emission during and outside periastron. Therefore the observation was divided into two different segments: the first exposure of $\sim110$\,ks (ObsID~0763720401) was performed on 2016~March~1 with the system out of periastron; the second exposure of $\sim110$\,ks (ObsID~0763720301) was performed on 2016 March 21 near the periastron passage. Neither the spatial nor spectral resolution of {\it XMM-Newton} allows to separate the two components. X-ray data were therefore analyzed for the whole binary system.

We analyzed X-ray data gathered with the three EPIC instruments (PN, M1, and M2). EPIC data were first processed using the SAS V15.0 standard tasks. X-ray source events were extracted from a circle with a radius of 37.5$\arcsec$ centered on the target position. A similar nearby region, free from other X-ray sources, was used to extract background events. We considered only events with energy ranging between 0.3 and 7.9\,keV. During the two observations some time intervals were affected by significantly high background levels. We therefore discarded these high-background intervals, slightly reducing the actual exposure times.

\begin{table*}
	\centering
	\caption{Log of observations. Columns 1--2 list the UT date and heliocentric Julian date (HJD) of each observation. Our RV measurements from the disentangling of the Stokes $I$ LSD profiles are reported in columns 3--4. The estimated uncertainty of RV measurements is 0.9 {\kms}. Column 5 contains the orbital phase, columns 6--7 respectively list the S/N of the reduced observed spectra per spectral pixel at $\lambda = 566$~nm and the S/N of the Stokes $V$ LSD profiles relative to the unpolarized continuum level.
	}
	\label{table:obs}
	\begin{tabular}{ccccccc} 
		\hline
		Date        & HJD           & RV$_\text{A}$       & RV$_\text{B}$   & Phase & S/N spectrum & S/N LSD       \\
        (UTC)       & ($2400000+$)    & {\kms}     & {\kms} & &&    \\
		\hline
        2014-02-20 &	56708.83860 &	73.02 &	-2.35 &	0.232 &	113 &	7257 \\
        2016-01-14 &	57401.79896 &	57.22 &	10.96 &	0.308 &	129 &	8438 \\
        2016-01-14 &	57401.96585 &	56.76 &	11.85 &	0.312 &	107 &	6720 \\
        2016-01-15 &	57402.72847 &	53.89 &	16.12 &	0.330 &	129 &	8156 \\
        2016-01-15 &	57402.95519 &	52.99 &	17.30 &	0.336 &	125 &	7853 \\
        2016-01-16 &	57403.77253 &	49.67 &	21.17 &	0.356 &	136 &	9026 \\
        2016-01-16 &	57403.95313 &	48.55 &	21.49 &	0.361 &	124 &	8344 \\
        2016-01-17 &	57404.79009 &	45.17 &	24.46 &	0.381 &	136 &	8957 \\
        2016-01-17 &	57404.95526 &	44.35 &	24.67 &	0.385 &	125 &	8181 \\
        2016-01-18 &	57405.76619 &	40.89 &	28.08 &	0.405 &	141 &	9604 \\
        2016-01-18 &	57405.88228 &	40.42 &	28.67 &	0.408 &	141 &	9374 \\
        2016-01-19 &	57406.86790 &	34.42 &	35.36 &	0.432 &	140 &	9414 \\
        2016-01-20 &	57407.79079 &	34.39 &	36.93 &	0.455 &	122 &	7815 \\
        2016-01-21 &	57408.90430 &	30.19 &	43.33 &	0.483 &	106 &	7043 \\
        2016-01-22 &	57409.80457 &	26.31 &	47.17 &	0.505 &	120 &	8101 \\
        2016-01-23 &	57410.88334 &	20.12 &	50.32 &	0.531 &	128 &	8559 \\
        2016-01-24 &	57411.83838 &	16.44 &	54.14 &	0.555 &	136 &	8785 \\
        2016-01-25 &	57412.86504 &	12.99 &	57.81 &	0.580 &	120 &	7774 \\
        2016-01-26 &	57413.84815 &	9.65 &	60.31 &	0.604 &	126 &	8357 \\
        2016-01-27 &	57414.88632 &	6.38 &	62.87 &	0.630 &	128 &	8079 \\
        2016-01-28 &	57415.88710 &	3.64 &	66.27 &	0.655 &	127 &	8178 \\
        2016-01-29 &	57416.84774 &	1.18 &	69.26 &	0.678 &	128 &	8725 \\
		\hline
	\end{tabular}
\end{table*}

\section{Least-squares deconvolution}
\label{section:LSD}
Cool stars generally exhibit very weak Zeeman polarization signals in individual spectral lines. Combining the signal from a large set of spectral lines is a widely used solution in order to obtain a high S/N mean line profile. Here we applied the least squares deconvolution technique (LSD, \citealt{donati++1997-libre-esprit, kochukhov2010A&A...524A...5K}) to compute high S/N mean Stokes $IV$ profiles for each observation. LSD effectively performs a weighted co-addition of a set of spectral lines, defined by a line mask, into a mean profile. To construct the line mask, we retrieved a line list from the \texttt{VALD3} database \citep{vald2015} using a \texttt{MARCS} stellar atmosphere \citep{marcs2008} with {\teff} = 4750~K and {\logg} = 4.0. After having removed spectral regions where tellurics or particularly strong and broad spectral lines were present, we obtained a line mask containing 5304 atomic lines deeper than $20\%$ of the continuum. The LSD Stokes $IV$ profiles were then computed with the LSD code described by \citet{kochukhov2010A&A...524A...5K}. The mean parameters used to scale the LSD profiles were the mean wavelength $\lambda_0 = 5643$~{\AA} and the mean Land\'e factor $z_0 = 1.221$. The LSD procedure provided a median S/N gain of $66$ with a median S/N of the LSD Stokes $V$ profile of $8263$.

The resulting Stokes $IV$ profiles are shown in Fig.~\ref{fig:lsd}. The circular polarization signatures are detected unambiguously in both components throughout the part of the orbit sampled by our observations (between orbital phases 0.232 and 0.678). The Stokes $I$ profiles show that the primary component has somewhat deeper spectral lines. As our data mostly samples orbital phases where the two components are not clearly separated, we do not study variation of the mean longitudinal magnetic field for each component.

\section{Spectrum disentangling and orbital solution}
\label{section:spectral-disentangling}
V1878~Ori is a double-lined spectroscopic binary for which time-dependent superposition of the component spectra greatly complicates interpretation of the observed spectroscopic time series. In order to isolate the contribution from individual stars, we used different spectral disentangling methods. This allowed us to (a) measure accurate radial velocities for each of the two components at each epoch, (b) derive an orbital solution for the binary system, (c) obtain individual disentangled stellar spectra for each of the two components.

In this work, we used the disentangling algorithm described in \citet{2010MNRAS.407.2383F} and subsequently used by e.g. \citet{kochukhov2018MNRAS.478.1749K,rosen2018A&A...613A..60R,kochukhov2019ApJ...873...69K}. First, we modelled our series of the Stokes $I$ LSD profiles to measure the RV variation of the two components. The algorithm retrieves a mean Stokes $I$ profile for each component as well as a set of RV mesurements. It assumes that, at each epoch, the observed Stokes $I$ LSD profile is a combination of two profiles which stay constant throughout the time series and are shifted by a variable radial velocity. Starting with an initial guess of RVs according to the orbital solution by \citet{covino2001}, we derived the mean profiles for each component as well as RV measurements for each observing epoch. Our measurements are listed in columns 3--4 of Table~\ref{table:obs}.

We then fitted the RV measurements to derive an orbital solution. We used both our own measurements, which sample around half of the orbit, and the data from \citet{covino2001}, which, though less precise, has the benefit of covering the entire orbit. We present our orbital solution in Table~\ref{table:orbital-solution} alongside the previously determined solution by \citet{covino2001}. The orbital parameters derived from the RV fit include the orbital period $P_\text{orb}$, the time of periastron passage $T_0$, the eccentricity $e$, the radial velocity of the mass centre of the system $\gamma$, the radial velocity semi-amplitudes $K_{A,B}$, and the longitude of periastron $\omega$. We also determined the minimum stellar masses $M_{A,B}\sin^3i$ and the projected semi-major axes $a_{A,B}\sin i$. The RV measurements and the best-fitting model are shown in Fig.~\ref{fig:orbit_fit}. From the scatter of our RV points around the model, we determined that our measurements have a typical precision of 0.9 {\kms}.

\begin{figure*}
	\includegraphics[width=1.5\columnwidth]{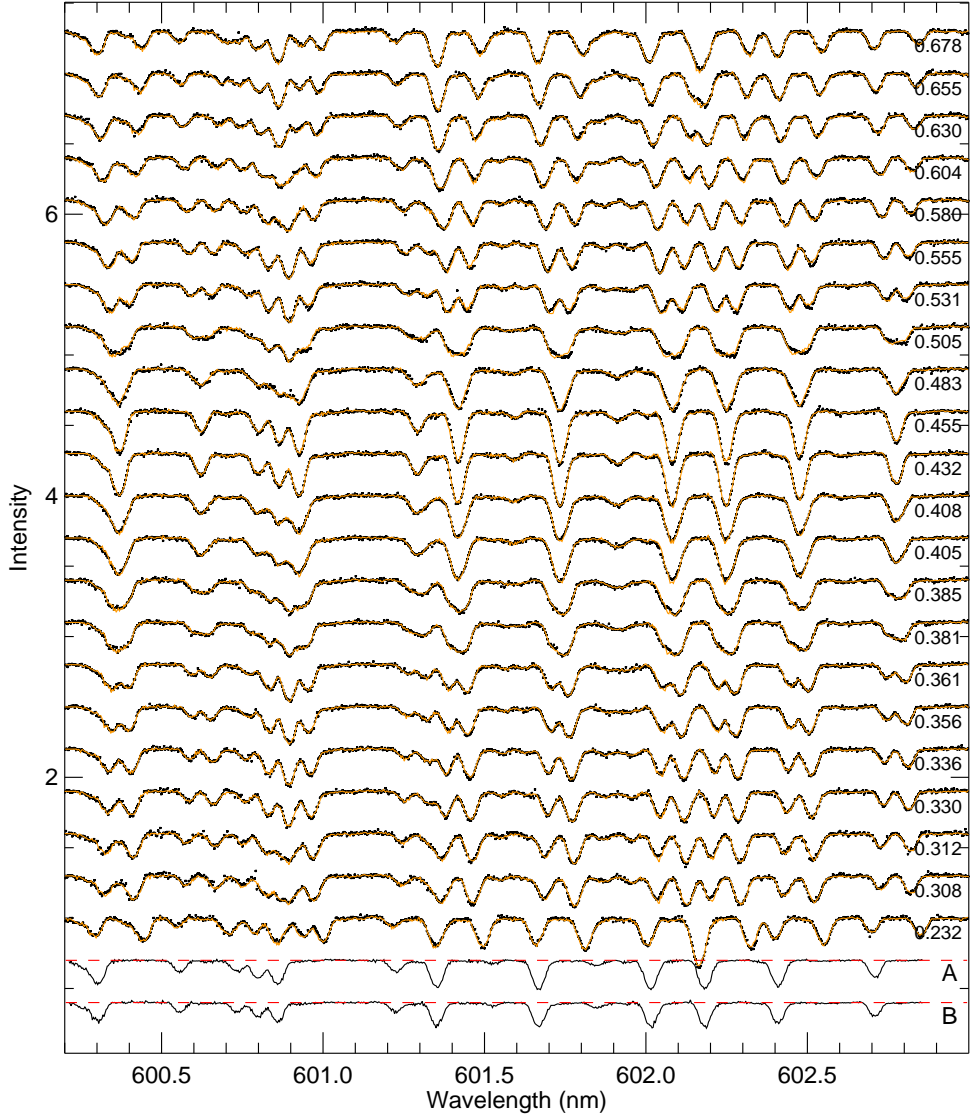}
  \caption{Illustration of the spectral disentangling. Observed spectra (black points) around 601~nm are shifted vertically according to the orbital phase (indicated on the right). The model spectrum at each phase is plotted with an orange line, and the resulting disentangled spectra for each components are plotted in the bottom of the figure with the continuum levels indicated with the red dashed lines.}
  \label{fig:sbfit_spec}
\end{figure*}

Finally, we applied the disentangling algorithm to the spectra themselves, working iteratively on 10--50~nm-long regions of the spectrum, to eventually cover a large part of the spectrum between 450 and 900~nm. In the disentangling procedure, we kept the previously found RV solution fixed, and recovered the individual spectra for each component, assuming that they exhibit no intrinsic variation. The disentangled spectra $S_{A,B}$ of each components were then rescaled in order to correct for the dilution of the continuum using the formula

\begin{equation}
\begin{array}{ll}
    S_A^0 = \left(S_A - \dfrac{1}{1+r_L}\right) \times \left(1+\dfrac{1}{r_L}\right) \\
    S_B^0 = \left(S_B - \dfrac{r_L}{1+r_L}\right) \times \left(1+r_L\right)
\end{array}
\end{equation}
where $r_L$ is the flux ratio of the two components, here $r_L = 1.0$. The spectra were finally individually renormalized by fitting the continuum level with a low-order polynomial. Fig.~\ref{fig:sbfit_spec} shows our set of observed spectra as well as the disentangled spectra for the A and B components in a narrow wavelength interval around 601~nm.

\begin{table}
	\centering
	\caption{Orbital parameters derived by \citet{covino2001} and in this work. }
	\label{table:orbital-solution}
	\begin{tabular}{lcc} 
		\hline
		        & \citet{covino2001}           & This work       \\
		\hline
        $P_\text{orb}$ (days)           & $40.5738 \pm 0.0047$	    & $40.58318 \pm 0.00025$	 \\
        $T_0$ (HJD)                     & $2451098.944 \pm 0.073$   & $2451098.925 \pm 0.065$\\
        $e$                             & $0.3186 \pm 0.0020$       & $0.3157 \pm 0.0018$\\
        $\gamma$ ({\kms})               & $+33.44 \pm 0.11$         & $+32.919 \pm 0.066 $\\
        $K_A$ (\kms)                    & $47.51 \pm 0.21$          & $47.08 \pm 0.18$ \\
        $K_B$ (\kms)                    & $47.59 \pm 0.15$          & $47.74 \pm 0.13$\\
        $\omega$ (deg)                  & $286.91 \pm 0.56$         & $287.45 \pm 0.51$ \\
        $M_A\text{sin}^3i$ $(M_\odot)$  & $1.541 \pm 0.014$         & $1.542 \pm 0.010$ \\
        $M_B\text{sin}^3i$ $(M_\odot)$  & $1.538 \pm 0.016$         & $1.520 \pm 0.013$\\
        $a_A\text{sin}i$ (Gm)           & $25.13 \pm 0.11$          & $24.930 \pm 0.097$\\
        $a_B\text{sin}i$ (Gm)           & $25.168 \pm 0.0083$       & $25.277 \pm 0.073$\\
		\hline
	\end{tabular}
\end{table}

\section{Stellar parameters}
\label{section:stellar-parameters}
We used the disentangled stellar spectra, corrected for the continuum dilution, to determine stellar parameters for the two components by fitting synthetic spectra to observations. To calculate theoretical stellar spectra and optimise free parameters, we used the Spectroscopy Made Easy (SME) spectrum synthesis code \citep{piskunov2017A&A...597A..16P}. Adopting the solar abundances from \citet{2009ARA&A..47..481A}, we downloaded absorption line lists from the {\tt VALD} database \citep{vald2015} using the {\tt Extract Stellar} mode with {\teff} = 4750~K and {\logg} = 4.0. We used 1D model atmospheres from the {\tt MARCS} grid \citep{marcs2008}. We fitted a set of wavelength intervals sensitive to different stellar parameters, mostly in the wavelength regions 516--519~nm and 600--620~nm, as used in \citet{valentifischer2005ApJS..159..141V}. The same fitting windows around lines of interest were used for both components.

\begin{figure*}
	\includegraphics[width=\textwidth]{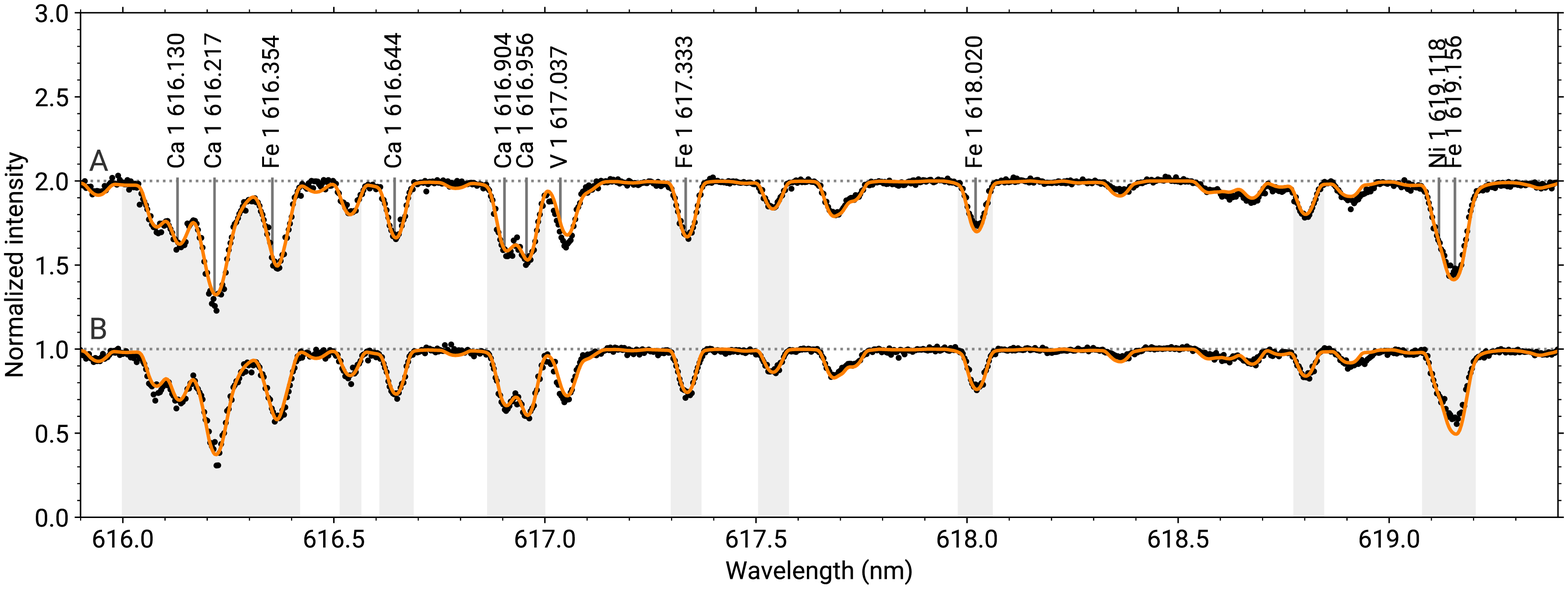}
    \caption{Disentangled observed spectra (black points) and best-fitting synthetic spectra (orange lines) for V1878~Ori A and B (the A component is shifted vertically). The intervals used for fitting the observed spectra are highlighted in light gray and a subset of the deepest lines is indicated with vertical lines.}
    \label{fig:sme}
\end{figure*}

We adopted a fixed macroturbulence {\vmac} = 0~{\kms} due to its degeneracy with {\vsini}. The instrumental broadening was modelled with a gaussian profile corresponding to a spectral resolution of $R = 65000$. The free parameters in the fit were the effective temperature {\teff}, the surface gravity {\logg}, the projected rotational velocity {\vsini}, the microturbulence {\vmic} and the overall metallicity [M/H]. The inferred stellar parameters are listed in Table~\ref{table:stellar-parameters}.  The error bars reported in this table were derived with the analysis of cumulative distribution functions, implemented in recent versions of SME \citep{piskunov2017A&A...597A..16P}.
Comparison of the best-fitting synthetic spectra and observations of the A and B components are shown in Fig.~\ref{fig:sme} for one of the wavelength regions used. Based on the CESAM PMS evolutionary tracks \citep{morel2008Ap&SS.316...61M,marques2013A&A...549A..74M} computed in  \citet[][]{villebrun2019A&A...622A..72V}, our best-fit stellar parameters correspond to masses of $1.7 \pm 0.3 M_\odot$ and $1.6 \pm 0.3 M_\odot$ for the primary and secondary, respectively and a young age of approximately 0.85 Myr. This suggests that both stars are beginning to develop radiative cores ($\sim 10\%$ by mass and $\sim 25\%$ by radius). It is important to note however, that there are large uncertainties in this region of the PMS HR diagram, and that another set of stellar evolutionary tracks such as the Yale-Potsdam Stellar Isochrones \citep[YaPSI;][]{spada2017ApJ...838..161S} would place our stars just across the fully-convective limit. Our spectra of V1878 Ori A and B confirm previous observations that these stars have an abnormally high Li abundance \citep{covino2001}, suggesting indeed a very young age. This however, is in contrast with the absence of high veiling or other signs of accretion onto the stars which would be expected at ages lower than 1~Myr. It is possible that binarity has impacted the accretion process very early in that case.

\begin{table}
	\centering
	\caption{Parameters of V1878~Ori A and B inferred with SME.}
	\label{table:stellar-parameters}
	\begin{tabular}{lcc} 
		\hline
		Component               & A           & B       \\
		\hline
		{\teff} (K)                 & $4800 \pm 126$   & $4759 \pm 200$\\
		{\logg} (dex)                & $4.07 \pm 0.42$  & $3.84 \pm 0.57$ \\
		{\vsini} (\kms)                & $15.4 \pm 2.2$   & $14.9 \pm 2.9$\\
		{[M/H]} (dex)                & $0.06 \pm 0.17$  & $-0.07 \pm 0.26$ \\
		{\vmic} (\kms)                 & $2.34 \pm 0.58$  & $1.29 \pm 0.77 $ \\
     		\hline
	\end{tabular}
\end{table}

\section{Zeeman Doppler imaging}
\label{section:zdi}
We carried out reconstruction of the large-scale magnetic field geometry of the two components simultaneously using the binary Zeeman Doppler imaging code {\tt InversLSDB} code described by \citet{rosen2018A&A...613A..60R}, which is based on the single star ZDI code {\tt InversLSD} code \citep{kochukhov2014A&A...565A..83K}. {\tt InversLSDB} has previously been used in the studies of the double-lined spectroscopic binary systems $\sigma^2$ CrB \citep{rosen2018A&A...613A..60R} and YY Gem \citep{kochukhov2019ApJ...873...69K}. Only the main dataset acquired in 2016 was used for ZDI inversions. The single observation from 2014 was discarded due to lack of precise rotation periods necessary for phasing together data taken two years apart and a high likelihood that magnetic field topologies have evolved during that time period.

In our analysis, we computed local Stokes $IV$ profiles using the Unno-Rachkovsky solution of the polarised radiative transfer equation, assuming a Milne-Eddington atmosphere. We also assumed that the LSD profiles behave as a single spectral line, with mean parameters, such as central wavelength and effective Land\'e factor, adopted from the LSD line mask applied to the observations. The local line profile strength was adjusted to reproduce the Stokes $I$ profiles. Additionally, the local profiles of the primary were scaled by a factor of $1.3$ to account for the difference in line depths between the primary and secondary.

\begin{figure}
	\includegraphics[width=0.9\columnwidth]{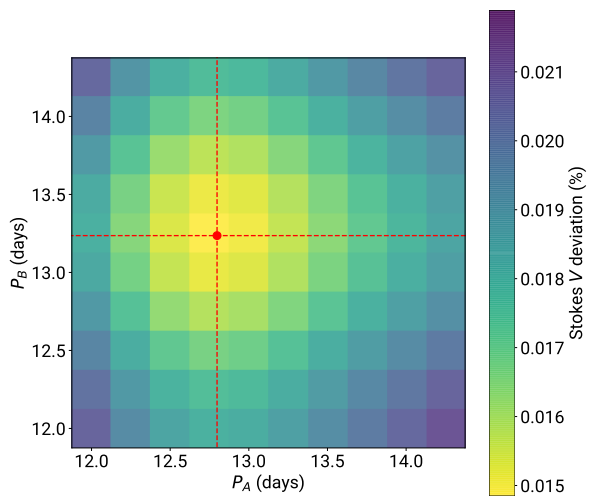}
    \caption{Deviation between observed and modelled Stokes $V$ LSD profiles as a function of the rotation periods $P_A$ and $P_B$ of the two components. The best-fitting period pair is shown with the red symbol and dashed lines.}
    \label{fig:P1vsP2}
\end{figure}

Moreover, we assumed that the stars have a spherical geometry and orbital parameters were taken from Table~\ref{table:orbital-solution}. As the V1878 Ori system has an eccentric and asynchronous orbit, it was necessary to adopt individual values of {\vsini}, rotation periods $P_{A,B}$, and inclination angles $i_{A,B}$ for each component in our analysis. We adopted the {\vsini} values from Table~\ref{table:stellar-parameters} that were determined using spectrum synthesis. The individual rotation periods of the primary and the secondary were found by running ZDI inversions for a grid of varying rotation periods for the two stars using a step of 0.25 days. For each inversion, we computed the deviation between the observed and reconstructed Stokes $V$ profiles, as illustrated in Fig.~\ref{fig:P1vsP2}. The best combination of periods, $P_A = 12.81$~d and $P_B = 13.24$~d, was determined by fitting a 2-D spline function to the deviation and finding the location of the minimum. Similarly, the inclination of the two stars was constrained -- more coarsely -- by running inversions for a grid of inclinations for the two stars with a step of 20 degrees. We adopted the best-fitting inclination values within that grid, namely $i_A = i_B = 60^\circ$. This inclination value, together with the rotation periods and \vsini\ that we adopted, imply stellar radii of $R_A = R_B = 4.5$~$R_\odot$. This is larger than $R_A = R_B = 3.4$--3.6~$R_\odot$ corresponding to the parameters obtained by \citet{covino2001} and \citet{marilly2007}. These inclination values imply stellar masses of $M_A \sim 2.4 M_{\odot}$ and $M_B \sim 2.3 M_{\odot}$ using the $M_{A,B}\text{sin}^3i$ values from Table~\ref{table:orbital-solution}. These are in disagreement with the stellar masses estimated from PMS evolutionary tracks in Sect.~\ref{section:stellar-parameters} ($M_A = 1.7 \pm 0.3 M_{\odot}$ and $M_B = 1.6 \pm 0.3 M_{\odot}$). While we believe the projected masses $M_{A,B}\text{sin}^3i$ infered from the RV solution to be accurate, the discrepancy can stem from the inclination values which are not well-constrained by the ZDI inversion or from the masses derived with the PMS models which can potentially suffer from systematic uncertainties.

The brightness mapping was carried out modelling the Stokes $I$ LSD profiles with our inversion code {\tt inversLSDB}. This reconstruction employed Tikhonov regularization, following the approach described in \citet{rosen2018A&A...613A..60R}, in order to obtain a solution with minimum brightness contrast.

The magnetic field of each star was parametrized as a combination of poloidal and toroidal fields, each decomposed using spherical harmonics as described in \citet{kochukhov2014A&A...565A..83K}. The spherical harmonic coefficients $\alpha_{\ell,m}$ and  $\beta_{\ell,m}$ represent respectively the contributions of the radial and horizontal poloidal components, and $\gamma_{\ell,m}$ the horizontal toroidal component. The parameters $\ell$ and $m$ are, respectively, the angular degree and the azimuthal order of the spherical harmonic function. Here, we set the maximum degree of the spherical harmonic coefficients to $\ell_{\rm max} = 10$. This choice of $\ell_{\rm max}$ does not restrict the complexity of the recovered field given the moderate \vsini of the studied stars, as we let this role be played by the regularization in our analysis. Indeed, a ZDI reconstruction using Stokes $V$ data alone is an ill-posed inverse problem, and one needs to use regularization in order to obtain a stable and unique solution. We adopted the following regularization function \citep{kochukhov2014A&A...565A..83K}
$$R = \Lambda \sum_{\ell=1}^{\ell_{\text{max}}} \sum_{m=-\ell}^{\ell} \ell^2 (\alpha_{\ell,m}^2 + \beta_{\ell,m}^2 + \gamma_{\ell,m}^2)$$ where $\Lambda$ is the regularization parameter. This function penalises the presence of high order modes, hence favouring simpler magnetic field topologies. The optimal value of the regularisation parameter $\Lambda$ was determined following \citet{kochukhov2017A&A...597A..58K}, by running inversions with a grid of successively decreasing regularization parameter, selecting the value that offers a good fit to the data without fitting the noise.

\begin{figure*}
	\includegraphics[width=1.9\columnwidth]{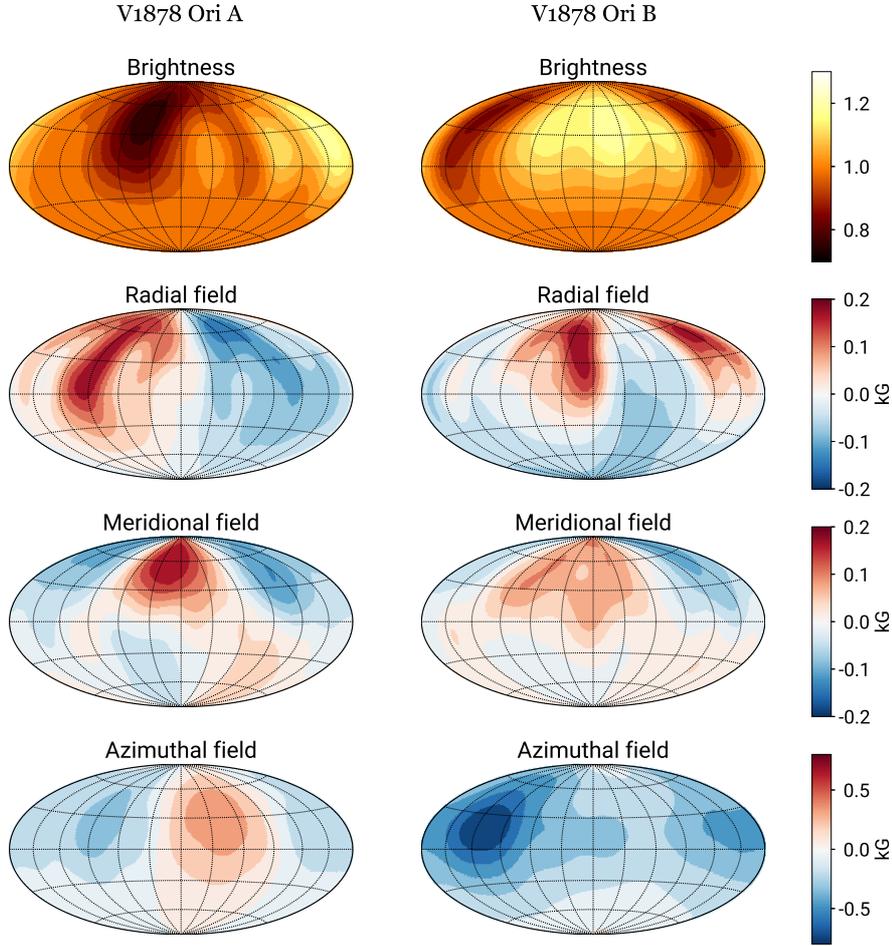}
    \caption{Global magnetic field maps of V1878 Ori A (left column) and V1878 Ori B (right column) displayed in the equal-area Hammer-Aitoff projection. The central meridian corresponds to $180^\circ$ longitude, or rotational phase 0.5, with the longitude increasing from left to right. The field strengths in kG is indicated by color bars on the right.}
    \label{fig:hammer}
\end{figure*}

\begin{figure}
	\includegraphics[width=1.0\columnwidth]{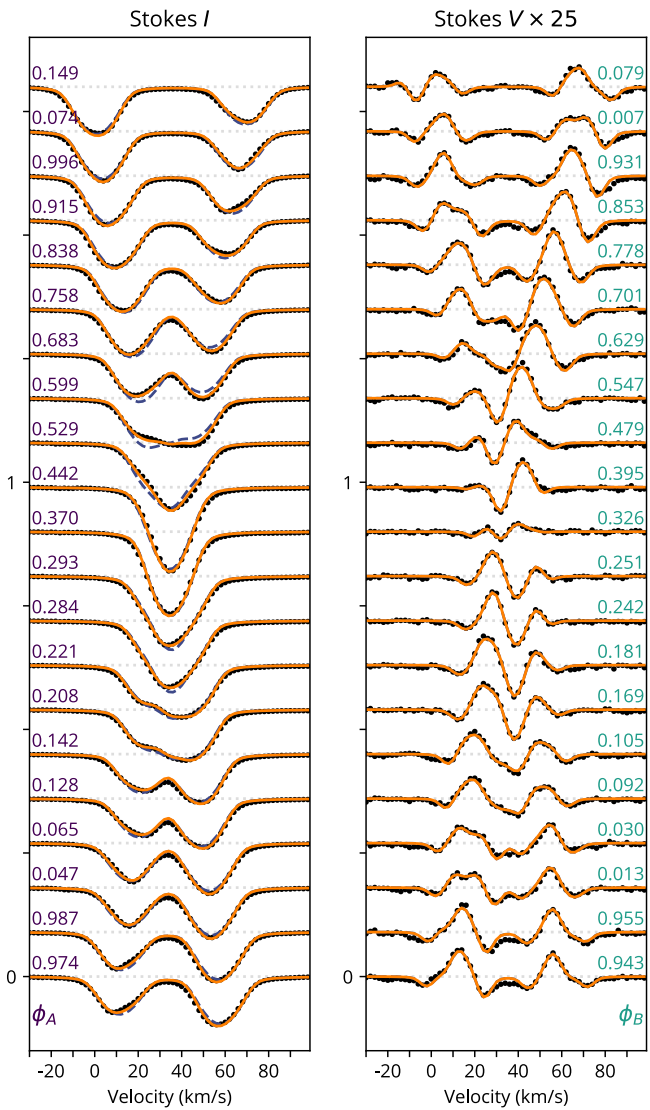}
    \caption{Observed (black symbols) and computed (solid orange lines) Stokes $IV$ LSD profiles shifted vertically with rotational phase rising upwards. The black dashed line in the Stokes $I$ panel shows the effect of removing brightness spots. The rotational phases of the primary and the secondary components are indicated on the left- and right-hand side of the plot, respectively.}
    \label{fig:zdi_prf}
\end{figure}

The recovered brightness and magnetic field maps are shown in Fig.~\ref{fig:hammer} while the comparison between observed and model LSD Stokes $IV$ profiles is presented in Fig.~\ref{fig:zdi_prf}.  The summary of the magnetic field characteristics for each component is gathered in Table~\ref{table:inversion_parameters}. The observed Stokes $I$ and $V$ LSD profiles are reproduced very well at all orbital phases.
The presence of spots explains the apparent reversal of the line depth ratio between the primary and secondary components, observed at the rotational phases $\phi_A$ between 0.529 and 0.683 (corresponding to the orbital phases 0.505--0.531 in Figs.~\ref{fig:lsd} and \ref{fig:sbfit_spec}). As indicated by the dashed line in Fig.~\ref{fig:zdi_prf}, this feature is unexplained if one assumes a homogeneous brightness distribution, resulting in a noticeably inferior line profile fit.

The large-scale magnetic fields of the two stars have rather different topologies. On the one hand, the magnetic field of the primary is mostly poloidal (80\% of the magnetic energy is concentrated in poloidal components) and non-axisymmetric (91\% in the modes with $|m|>\ell/2$). On the other hand, the field of the secondary is mostly toroidal (87\% of the field energy is in the toroidal modes) and axisymmetric (87\%). Additionaly, there is also a large difference in the average magnetic field strength of the two components. The primary has the weaker global magnetic field of the pair, with a mean field strength of 180~G and a maximum strength of 410~G. The secondary has a roughly twice stronger field, with a mean field strength of 320~G and a maximum strength of 810~G. Fig.~\ref{fig:hammer} suggests that this difference is due to a strong axisymmetric toroidal field, dominating the azimuthal field map of the secondary but absent in the primary.
As the two components have similar stellar parameters and rotation periods, it could be expected -- if a dynamo mechanism generating their magnetic field was strictly a function of the stellar parameters -- that their magnetic field would be similar. But here, as already observed before in fully-convective stars \citep{kochukhov2017ApJ...835L...4K} and rapidly rotating Sun-like stars \citep{rosen2018A&A...613A..60R}, two components of a binary system with similar stellar parameters exhibit significantly different field configurations. However, a trivial interpretation of this observation is that one or both of the stars have variable magnetic activity and the observations analysed here correspond to one particular phase of activity cycle, not necessarily representative of the time-averaged activity levels of the two components.

\begin{table}
	\centering
	\caption{Summary of the global magnetic field characteristics of V1878 Ori A and B.}
	\label{table:inversion_parameters}
	\begin{tabular}{rcc} 
		\hline
        Distribution of the		&		&		\\
		magnetic field energy 	& V1878 Ori A	&V1878 Ori B	\\
		\hline
		$\ell=1$	                    & 70.0 \%	& 89.4 \% \\
        $\ell=2$	                    & 23.0 \%	& 7.6  \% \\
        $\ell=3$	                    & 4.1  \%	& 1.7  \% \\
        $\ell=4$ 	                    & 1.3  \%	& 0.4  \% \\
        $\ell=5$	                    & 0.7  \%	& 0.4  \% \\
        $\ell=6$	                    & 0.5  \%	& 0.3  \% \\
        $\ell=7$	                    & 0.3  \%	& 0.1  \% \\
        $\ell=8$	                    & 0.1  \%	& 0.0  \% \\
        $\ell=9$	                    & 0.0  \%	& 0.0  \% \\
        $\ell=10$                       & 0.0  \% 	& 0.0  \% \\
		poloidal	                    & 79.7 \%	& 12.8 \% \\
		toroidal                        & 20.3 \%   & 87.2 \% \\
		axisymmetric ($|m|<\ell/2$) 	& 8.5  \%	& 86.5 \% \\
		\hline
		Magnetic field strength & (G)& (G) \\
		\hline
		$\bb$							&	180     & 310       \\
		$|B|_\mathrm{max}$			    &	410     & 810		\\
		\hline
	\end{tabular}
\end{table}

\section{X-ray data analysis}
\label{section:xraydata}

\begin{figure*}
    \centering
    \includegraphics[width=1.9\columnwidth]{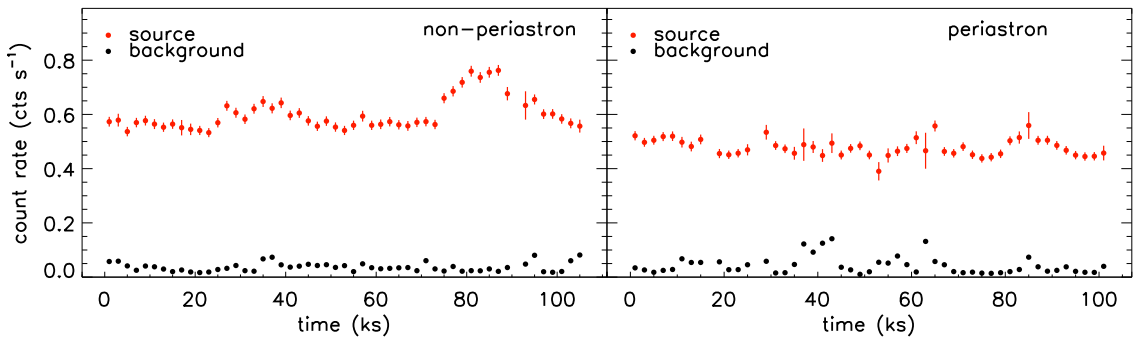}
    \caption{Background-subtracted X-ray light curves of V1878~Ori  observed during non-periastron and periastron observing segments. Count rates were obtained by adding the photons registered by the three EPIC instruments and adopting a time bin of 2\,ks. Error bars correspond to $1\sigma$.}
    \label{fig:xraylightcurve}
\end{figure*}

To study the X-ray emission of V1878~Ori observed by {\it XMM-Newton} during non-periastron (companion separation $\sim21~R_\star$) and periastron (companion separation $\sim11~R_\star$) we first inspected the X-ray light curves, obtained by adding all the events registered by the three EPIC instruments (Fig.~\ref{fig:xraylightcurve}). During both segments, the X-ray emission of V1878~Ori shows significant variability, likely due to constant flaring activity. For the strongest flare, occurring during the final 30\,ks of the non-periastron segment, the X-ray count rate increases up to 30\%. In addition to this short timescale variability, the average X-ray emission level during the two observing intervals, separated by $\sim20$\,d, was different: during the non-periastron phase the average X-ray count rate was $0.601\pm0.003$\,cts\,s$^{-1}$, significantly higher than $0.477\pm0.003$\,cts\,s$^{-1}$, observed during the periastron.

To derive the characteristics of the coronal plasma we inspected the PN, M1, and M2 X-ray spectra collected during non-periastron and periastron, respectively. To increase the signal-to-noise ratio we rebinned each spectra to obtain at least 30 counts in each channel. To better constrain the X-ray emitting plasma properties we simultaneously fitted the spectra of the three EPIC instruments. The spectral analysis was performed using XSPEC V12.10. We adopted a model of an optically-thin plasma emission \citep[using the APED atomic database,][]{2001ApJ...556L..91S} with two isothermal components, and taking into account also the interstellar absorption. We left as free parameters the abundances of O, Ne, S, and Fe, while the abundances of the other elements were linked to the iron one.

\begin{figure*}
    \centering
    \includegraphics[width=1.9\columnwidth]{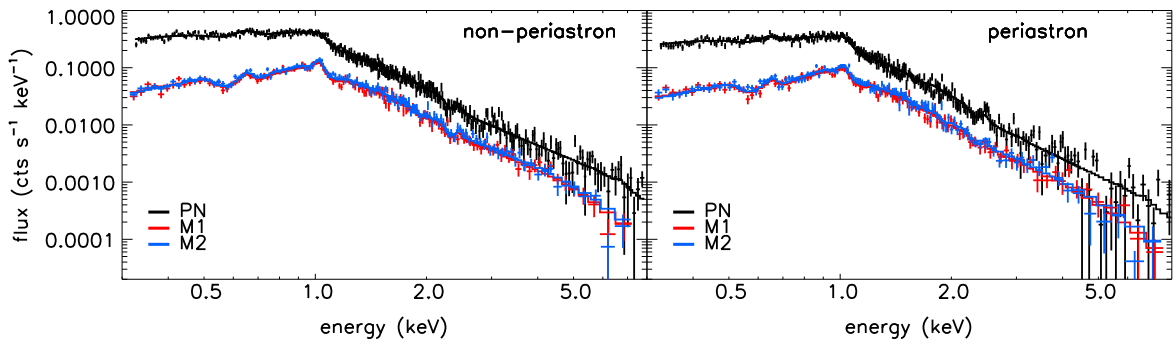}
    \caption{X-ray spectra of V1878~Ori observed during non-periastron and periastron observing segments. Dots with error bars indicate the observed spectra (black, red, and blue mark PN, M1, and M2, respectively), continuous histograms represent predicted spectra. Error bars correspond to $1\sigma$.}
    \label{fig:xrayspec}
\end{figure*}

The observed and predicted spectra are shown in Fig.~\ref{fig:xrayspec}. The values of the best-fit parameters are reported in Table~\ref{table:xrayparam}. In both of the observing segments the temperature of the hot component is unconstrained, and only a lower limit could be obtained. The plasma properties during the two segments are very similar, with the non-periastron X-ray emission showing only a slightly larger amount of plasma emission measure ($EM$) in both the thermal components. From the models that we derived for V1878~Ori, we obtain X-ray luminosities of $1.6\times10^{31}$\,erg\,s$^{-1}$ and $1.2\times10^{31}$\,erg\,s$^{-1}$ in the $0.5-8.0$\,keV band, for the non-periastron and periastron phases, respectively.

A hot and X-ray bright corona, such as that of V1878~Ori, is typical for young late-type stars \citep[e.g.][]{2005ApJS..160..401P}. In particular, V1878~Ori displays average plasma temperatures, luminosity, and abundances comparable to those of other young intermediate-mass stars with convective envelopes \citep[e.g.][]{2017A&A...598A...8P,2009ApJ...703..252J}.

The general picture of coronal physics of late type stars, which implies that hot coronal plasma is heated and confined by stellar magnetic fields, is robust. However some fundamental aspects, e.g. the extent of the magnetic structure, are still debated. Monitoring the coronal emission of eccentric binaries offers the opportunity to probe the magnetospheric extents, by investigating whether and how the two magnetospheres interact. Constraining and understanding the possible magnetospheric interactions in binary stars is also important for the understanding of star-planet magnetic interactions \citep[e.g.][]{2009ApJ...704L..85C,2010MNRAS.406..409F,2019AN....340..329P}.

Evidence of enhanced X-ray emission at periastron in eccentric binaries, suggesting interacting magnetospheres, was obtained for DQ~Tau \citep{2011ApJ...730....6G} and, tentatively, for a few other stars \citep{getman2016AJ....152..188G}. In that study, X-ray snapshot observations indicated a 50\% increase in the periastron flux relative to outside periastron but as only one non-periastron epoch was observed robust statistics could not be ascertained.

The two long X-ray observations of V1878~Ori obtained at and outside periastron indicate that, for this target and during this epoch, periastron passage was not associated with enhanced X-ray emission. Therefore the magnetospheric interaction of the two companions at periastron was not strong enough to affect the coronal activity. This lack of magnetospheric interaction in V1878~Ori, if compared to the previous results obtained for DQ~Tau, may be explained by the smaller periastron separation of DQ~Tau with respect to V1878~Ori ($8\,R_{\star}$ vs. $11\,R_{\star}$), different field geometries, and, possibly, by an even stronger magnetic field in the cooler young low-mass component stars of DQ~Tau, although these have not yet been measured for DQ Tau. The upper limit inferred for the coronal extent of V1878~Ori agrees with the value of $\sim2\,R_{\star}$ inferred for the coronal structure dimension, obtained studying the supersaturation effects of coronal activity in young intermediate-mass stars \citep{2016A&A...589A.113A}.

\begin{table*}
    \centering
    \caption{Best fit parameters of the X-ray spectra of V1878~Ori collected during non-periastron and periastron phases. The plasma emission measure $EM$ is defined as $n_{\rm e}n_{\rm H}V$, where $n_{\rm e}$ and $n_{\rm H}$ are the electron and hydrogen densities, and $V$ is the plasma volume. Abundances refer to the solar photospheric values of \citet{1989GeCoA..53..197A}. Uncertainties correspond to 1$\sigma$.}
    \label{table:xrayparam}
    \begin{tabular}{ccccccccccc}
        \hline
            & $N_{\rm H}$            & $T_1$ & $T_2$ & $\log EM_1$  & $\log EM_2$ & O             & Ne             & S             & Fe \\
        obs  & (10$^{20}$\,cm$^{-2}$) & (MK)  & (MK)  & (cm$^{-3}$)  & (cm$^{-3}$) & (O$_{\odot}$) & (Ne$_{\odot}$) & (S$_{\odot}$) & (Fe$_{\odot}$) \\
        \hline
        non per.    & $2\pm1$   & 11.2$^{+0.7}_{-1.2}$  & $>38$     & 54.28$^{+0.08}_{-0.12}$   & 53.52$^{+0.24}_{-0.16}$ & 0.35$^{+0.15}_{-0.12}$ & 0.7$^{+0.3}_{-0.2}$ & 0.21$\pm0.16$ & 0.04$\pm0.02$ \\
        per.  & $<3$    & 11.5$^{+0.6}_{-0.9}$ & $>12$ & 54.20$^{+0.07}_{-0.13}$ & 53.2$^{+0.9}_{-0.2}$ & 0.35$^{+0.3}_{-0.14}$  & 0.6$^{+0.7}_{-0.3}$ & 0.21$\pm0.18$ & 0.05$^{+0.08}_{-0.02} $\\
        \hline
    \end{tabular}
\end{table*}

\section{Summary and discussion}
\label{section:results+discussion}
This work presented the spectropolarimetric analysis of the double-lined spectroscopic binary system V1878 Ori, consisting of two intermediate-mass weak-line T Tauri stars. We used time-resolved high-resolution spectropolarimetric observations to derive high signal-to-noise ratio LSD profiles, measure the radial velocities of the two components and refine the system's orbital solution. We obtained disentangled intensity spectra for the two stars and inferred stellar parameters with spectrum synthesis modelling. Finally, we applied the ZDI technique to the time-series of composite Stokes $I$ and $V$ LSD profiles to simultaneously reconstruct the brightness as well as vector magnetic field maps of V1878 Ori A and B.

We found that the secondary has a stronger global field than the primary. These fields also turned out to have different topological properties. The global magnetic geometry of the primary is predominantly poloidal and non-axisymmetric while the field of the secondary is mostly toroidal and axisymmetric. Individually, both stars agree well with the trend that strong toroidal fields appear predominantly axisymmetric, as found by \citet{see2015MNRAS.453.4301S} based on ensemble analysis of single-star ZDI studies.

This is the first ZDI study of a binary system consisting of two intermediate-mass T Tauri stars. The overall properties of the reconstructed magnetic fields are in broad agreement with previous studies of single T~Tauri stars: the mean large-scale magnetic field (i.e. recovered from spectropolarimetric measurements subject to flux cancelation) is stronger than for hotter fully-radiative IMTTS on which the surface magnetic field is too weak and/or too complex to be detected \cite[][]{villebrun2019A&A...622A..72V} but weaker than the simple strong axisymmetric fields recovered in cool fully convective  T~Tauri stars \cite{hill2019}. Also in agreement with these studies, for \teff$>4300$~K the large-scale surface magnetic topology is more complex than the simple superposition of a dipole plus an octupole. A previous ZDI study of two higher-mass intermediate-mass T Tauri stars in the Chamaeleon cluster revealed even more complex large-scale magnetic field distributions than those recovered for V1878 Ori. As those stars were similar masses but more evolved than the V1878 Ori primary and secondary stars, they have much larger radiative cores ($>60\% R_\odot$; this indicates that the field complexity in V1878 Ori will likely also increase as its radiative core developes further \citep{hussain2009MNRAS.398..189H}. This study hints that the magnetic fields of intermediate-mass stars can change a lot around the convective boundary, but we have relatively few magnetic studies of stars in this region \citep[see][]{villebrun2019A&A...622A..72V} thus far. More magnetic studies of candidates covering this region of the PMS are needed to better understand the evolution of the stellar dynamo in young stars. 

The fact that the two V1878 Ori stars, which have nearly identical stellar parameters and the same evolutionary history, have radically different magnetic field characteristics is striking.
For fully convective stars, different magnetic fields at similar stellar parameters (as observed in e.g. GJ65 A and B, \citealt{kochukhov2017ApJ...835L...4K}) can be tentatively explained by a bi-stability of the dynamo process \citep{gastine2013A&A...549L...5G}. For partially convective binary stars showing the same phenomenon, such as $\sigma^2$~CrB \citep{rosen2018A&A...613A..60R} and V1878 Ori, such an explanation does not hold. Rather, our results may provide a clue that the spread in the ratio of toroidal to poloidal magnetic energy of partially convective stars visible in Figure~2 of \cite{see2015MNRAS.453.4301S} is intrinsic to dynamo action in these stars and not related to a hidden parameter (such as age, composition, or initial conditions). Here, the only parameter that appeared significantly different between the two components of V1878 Ori in the spectrum synthesis is the microturbulence.

We now know that young rapidly rotating partially convective stars can exhibit a wide range of toroidal magnetic energy fractions \cite{folsom2016MNRAS.457..580F,folsom2018MNRAS.474.4956F}. Future observations monitoring systems such as V1878 Ori for a period of years to decades will allow us to devise to which extent the differences observed in the magnetic topologies of V1878~Ori A and B can be attributed to intrinsic dynamo-related variability, whether cyclic or chaotic. Indeed, for main sequence partially convective stars, the fraction of toroidal magnetic energy as well as the degree of axisymmetry of the poloidal component can display dramatic variations along the magnetic cycle \citep{dunstone2008MNRAS.387.1525D,boro-saikia2018A&A...620L..11B}.

It is also important to note that the strength of the field recovered using ZDI is a small fraction of the total magnetic field strength at the surface of the star \citep{kochukhov2019ApJ...873...69K,lavail2019A&A...630A..99L,kochukhov2020}. This fraction depends on the magnetic field complexity, as (potentially strong) local tangled fields can cancel out in circular polarization. Analysis of Zeeman broadening in intensity spectra would be needed to obtain a reliable estimate of the total magnetic field strength but was not carried out with this dataset as Zeeman broadening is relatively small in the optical.
Self-consistent Zeeman broadening measurements and ZDI inversions, as well as spot coverage measurements using molecular bands are becoming more feasible using high-resolution near-infrared spectropolarimeters such as SPIRou at CFHT or the upcoming CRIRES+ instrument at VLT. These will certainly help to obtain a more complete picture of the magnetism of these objects and understanding of the dynamo processes taking place in the interiors of pre-main sequence stars such as these.

\section*{Acknowledgements}
\label{section:acknowledgements}
O.K. acknowledges support by the Knut and Alice Wallenberg Foundation (project grant `The New Milky Way'), the Swedish Research Council (projects 621-2014-5720, 2019-03548), and the Swedish National Space Board (projects 185/14, 137/17). E.A. and the BinaMIcS Collaboration acknowledge financial support from `Programme National de Physique Stellaire' (PNPS) of CNRS/INSU (France).
This research has made use of NASA's Astrophysics Data System.
Based on observations obtained at the Canada-France-Hawaii Telescope (CFHT) which is operated from the summit of Maunakea by the National Research Council of Canada, the Institut National des Sciences de l'Univers of the Centre National de la Recherche Scientifique of France, and the University of Hawaii. The observations at the Canada-France-Hawaii Telescope were performed with care and respect from the summit of Maunakea which is a significant cultural and historic site.

\section*{Data availability}
\label{section:data-availability}
The spectropolarimetric observations analysed in this article are available from the PolarBase database \citep{petit2014PASP..126..469P} at \url{http://polarbase.irap.omp.eu}. The XMM-Newton data are available from \url{http://nxsa.esac.esa.int/nxsa-web/\#search}.
%



\bibliographystyle{mnras}
\bibliography{article} 







\bsp	
\label{lastpage}
\end{document}